\documentclass[aps,prl,twocolumn,superscriptaddress,showpacs]{revtex4}

\usepackage{graphicx}
\usepackage{dcolumn}
\usepackage{bm}
\begin{document}

\title{Parametric frequency mixing in the magneto-elastically driven FMR-oscillator}


\author{C.L. Chang}
\affiliation{Zernike Institute for Advanced Materials, University
of Groningen, Groningen, The Netherlands}
\author{A.M. Lomonosov}
\affiliation{LAUM CNRS 6613, Universit\'e du Maine, 72085 Le Mans
cedex, France}
\author{J. Janusonis}
\affiliation{Zernike Institute for Advanced Materials, University
of Groningen, Groningen, The Netherlands}
\author{V.S. Vlasov}
\affiliation{IMMM CNRS 6283, Universit\'e du Maine, 72085 Le Mans
cedex, France} \affiliation{Syktyvkar State University named after
Pitirim Sorokin, 167001, Syktyvkar, Russia}
\author{V.V. Temnov}
\email{vasily.temnov@univ-lemans.fr} \affiliation{IMMM CNRS 6283,
Universit\'e du Maine, 72085 Le Mans cedex, France}
\affiliation{Fritz-Haber-Institut der Max-Planck-Gesellschaft,
Abteilung Physikalische Chemie, Faradayweg 4-6, 14195 Berlin,
Germany}
\author{R.I. Tobey}
\email{r.i.tobey@rug.nl} \affiliation{Zernike Institute for
Advanced Materials, University of Groningen, Groningen, The
Netherlands}

\begin{abstract}
We demonstrate the nonlinear frequency  conversion of
ferromagnetic resonance (FMR) frequency  by optically excited
elastic waves in a thin metallic film on dielectric substrates.
Time-resolved probing of the magnetization directly witnesses
magneto-elastically driven second harmonic generation, sum- and
difference frequency mixing from two distinct frequencies, as well
as parametric downconversion of each individual drive frequency.
Starting from the Landau-Lifshitz-Gilbert equations, we derive an
analytical equation of an elastically driven nonlinear parametric
oscillator and show that frequency mixing is dominated by the
parametric modulation of FMR frequency.
\end{abstract}

\maketitle

Parametric behaviour emerges in a wide range of periodically
driven systems when their parameters  are also periodically
modulated\cite{Butikov}. Examples can be found in
nano-optomechanical \cite{Rugar, Aspelmeyer, Papariello2016} and
micro-electromechanical systems \cite{ZhangSA2002},  (spin) wave
dynamics\cite{L'vov}, quantum circuitry\cite{Castellanos-Beltran},
energy harvesting applications\cite{JiaSciRep2016}, and in line
with our current report, magneto-mechanical
systems\cite{Zhange1501286} including spin pumping
capabilities\cite{KeshtgarSSC2014}.  The utility of parametric
behaviour has been shown for quantum limited detection, noise
floor reduction or low noise amplification of small
signals\cite{Castellanos-Beltran,Rugar}.

Parametric phenomena in magnetization dynamics have also been
extensively studied in the framework of spintronic and magnonic
applications\cite{Serga}, where the  downconversion of a
microwave-driven uniform precession can generate two
counter-propagating spin waves of varying frequency and
wavevector.  The onset of parametric behaviour in these cases is
monitored via the enhanced damping and linewidth changes of the
ferromagnetic resonance (FMR) precessional motion. Furthermore,
time domain probing of FMR precession modulated with multiple
microwave electromagnetic fields leads to seeded parametric
downconversion\cite{ElezzabiAPL2003}. Additional studies along
these lines have resulted in the generation and detection of a
range of frequency mixing processes of both uniform precessional
modes as well as higher energy spin waves\cite{CapuaPRL2016,
GerritsPRL2007, Bauer}, including frequency up- and
down-conversion.

Looking beyond microwave excitation, the overlapping frequency
range of (surface) acoustic waves and magnetization precession
provides for a unique opportunity to study their interactions and
to explore physical processes where coherent elastic deformations
could provide the necessary parametric modulation to drive complex
magnetization dynamics.  In recent years, magnetoelastic
interactions have seen a resurgence of interest, and linear
coupling between these degrees of freedom have been demonstrated
\cite{Weiler2, Thevenard2, Dreher, Kim, Scherbakov, Jaeger1}.  To
our knowledge, only a single report has discussed the potential
for nonlinearities in the magnetoelastic
interactions\cite{Afanasiev}.

In this report we present experimental evidence for the nonlinear
(in the sense of frequency mixing) interaction between multiple
coherent elastic deformations and the magnetization precession in
a thin ferromagnetic film.  To explain our results we perform
analytical calculations of the Landau-Lifshitz-Gilbert equation,
subject to the periodic excitation of a large amplitude, coherent
elastic wave and show that the resulting dynamics can be described
by an extended Mathieu equation for a nonlinear parametric
oscillator.  Simulation results based on our theory show excellent
correspondence with experimental results and allow us to identify
a range of upconversion responses (enumerated below) as well as
the downconverted precessional response commonly associated with
parametric modulation.

The enabling feature of the present research is our recent
demonstration of a simple optical technique that is able to
generate multiple elastic waves utilizing the all-optical
ultrafast transient grating (TG) technique\cite{Janusonis2016_1},
which facilitates the excitation and detection of multiple
distinct elastic perturbations that propagate along the surface of
a magnetic thin film. In our recent reports, we have identified
them as the Rayleigh Surface Acoustic Wave (SAW) and the Surface
Skimming Longitudinal Wave (SSLW), oscillating at distinct
frequencies $\omega_{SAW}$ and $\omega_{SSLW}$. These acoustic
transients interact simultaneously with the FMR precession in a
nickel film. Varying the applied magnetic field and underlying
substrate material allows us to engineer the frequencies and
relative elastic excitation strength of the two waves, to
experimentally observe sum and difference frequency generation
(SHG and DFG at $\omega_{\pm}=\omega_{SSLW}\pm\omega_{SAW}$,
respectively) and second harmonic generation (SHG, both for
$\omega_{SHG}=\omega_{SAW}+\omega_{SAW}$ and
$\omega_{SHG}=\omega_{SSLW}+\omega_{SSLW}$).

In the TG geometry the sample is excited by two, spatially and
temporally coincident optical pulses generating a spatially
periodic, instantaneous excitation.  Due to thermoelastic
mechanisms, large amplitude elastic waves propagate along the
surface of the film (as well as into the bulk of the film).  A
time delayed probe measures the magnetization of the sample based
on the rotation of the polarization of the transmitted beam
(Faraday configuration), and is sensitive to the out-of-plane
component of the magnetization as it precesses under the action of
the elastic waves.

\begin{figure}
    \centering
    \includegraphics[width=0.9\columnwidth]{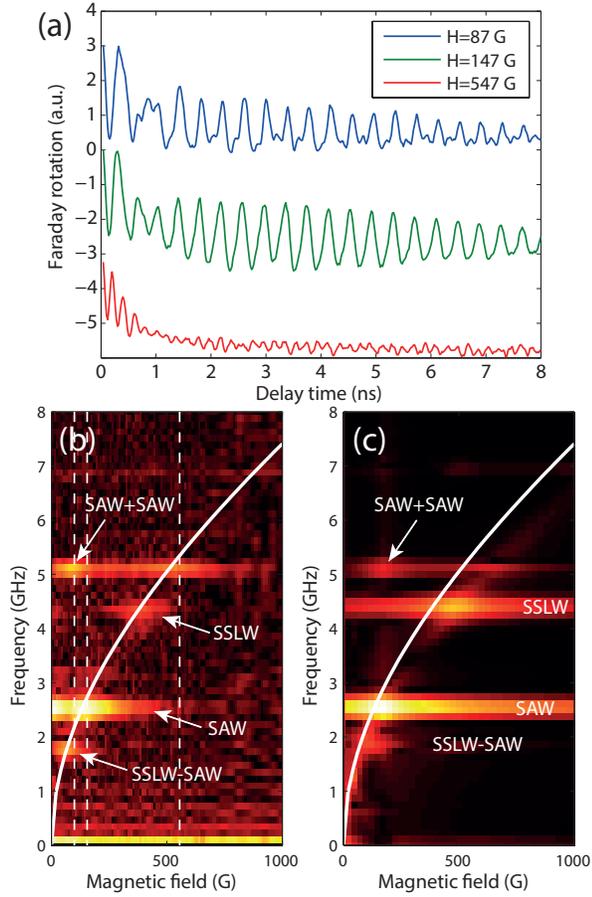}
    \caption{
    (a) The time-resolved Faraday rotation for three representative applied fields, exhibiting a range of magnetoelastic responses. (b) The Fourier spectra of Faraday time traces vary strongly with the applied magnetic field, and show maxima when resonant with the elastic driving fields (linear response) as well as at their sum and difference frequencies. Magneto-elastic frequency mixing is observed at the second harmonic frequency of the Rayleigh Surface Acoustic Wave (SAW) and difference frequency with a weaker Surface Skimming Longitudinal Wave (SSLW) wave, (c) The numerical solution of the parametric oscillator equation (Eq.~(1)) evidences the same behavior. White solid line marks the FMR-frequency $\Omega_0(H)/2\pi$, while vertical dashed lines identify the time traces in (a).}
    \label{fig:experimental}
\end{figure}

As an initial demonstration of magnetoelastic nonlinearities, we
revisit the Ni/MgO sample configuration we first described in
Janusonis et.al. \cite{Janusonis2015}.  On this substrate, the TG
excitation leads to a strong Rayleigh Surface Acoustic Wave
(Rayleigh SAW) and a barely - visible Surface Skimming
Longitudinal Wave (SSLW, see e.g.\cite{Janusonis2016}). Figure
1(a) shows the dependence of the magnetization precession
amplitude (Faraday rotation, vertically offset) on the magnitude
$H$ of the in-plane applied external magnetic field tilted by
angle $\phi=30^{\circ}$ with respect to TG wave vector. The
magnetic field allows for tuning the thin - film FMR frequency
following the Kittel formula
$\Omega_0=\gamma\mu_0\sqrt{H(H+M_0)}$, where $M_0$ is the
saturation magnetization in nickel and $\gamma$ is the
gyromagnetic ratio. For ease of visualization, the spectral
amplitude of Fourier transforms of individual scans taken over the
entire range of magnetic field are displayed in a 2D map as shown
in figure 1(b).  Within the range of magnetic field up to 1000~G,
the FMR frequency can be tuned to the underlying elastic
frequencies (2.55~GHz for SAW and weakly at 4.35~GHz for SSLW),
their difference (1.8~GHz) and sum (6.9~GHz) frequencies as well
as SAW second harmonic frequency (5.10~GHz). The elastic
frequencies are determined by the underlying substrate material
and are fixed once a TG period $\Lambda$ is experimentally
selected and shown here for the case of $\Lambda=2.2~\mu$m.  For
this combination of metal and substrate, elastic amplitude of SAW
is far larger than that of SSLW, resulting in a large precession
signal for the linear response at 2.55~GHz and its second harmonic
(5.10~GHz), and far weaker sum and difference frequency mixing
signals with the weak SSLW. We therefore consider this as a
monochromatic elastic wave (SAW) with a small additional
contribution of the SSLW.

Theoretical analysis of elastically driven Landau-Lifshitz-Gilbert
equations \cite{Kovalenko} shows that a relatively moderate value
of magnetostriction coefficient $b_1=1.5\times10^5$~J/m$^3$ in
nickel results in a small-angle FMR precession around the external
magnetic field (applied in the $xy$ - plane). The linearization of
LLG equations in the vicinity of the equilibrium magnetization
direction (see the Supplemantal Material for derivation) leads to
an equation of a driven parametric oscillator
\begin{equation}
\frac{d^2m}{dt^2} + \Gamma_0\frac{dm}{dt}+
(\Omega_0^2+\Omega_1^2e_{xx}(t))m=F_0e_{xx}(t)
\end{equation}
for the in-plane component $m=M_y(t)/M_0$ of the time-dependent
magnetization vector $\overrightarrow{M}(t)$. The dominant term
$e_{xx}(t)$ of the elastic strain represents a sum of two
contributions: a large amplitude, time periodic, SAW excitation at
frequency $\omega_{SAW}$  and a rapidly decaying SSLW excitation
in line with our previous Green's function calculation of the
elastic response\cite{Janusonis2016}.

Equation Eq.~(1) represents an approximation of a more complicated
equation (Eq.~(4) in the Supplemental Material). A detailed
analysis shows that that the damping term
$\Gamma(t)=\Gamma_0+\Gamma_1e_{xx}(t)$ is modulated by the elastic
strain $e_{xx}(t)$ as well and that there exist high-order
nonlinear terms proportional to $m^2e_{xx}$ and
$m\frac{dm}{dt}e_{xx}$. However, the dominant term in the sense of
frequency mixing are the parametric modulation of FMR-frequency
$\Omega^2(t)=\Omega_0^2+\Omega_1^2e_{xx}(t)$ and the external
driving force $F_0e_{xx}(t)$. An intrinsic property of our
methodology is that the parametric modulation
$\Omega^2_1(H,\phi)e_{xx}(t)=\frac{\gamma^2\mu_0b_1}{M_0}(H+M_0-[3H+2M_0]\cos^2\phi)e_{xx}(t)$
and the external driving force
$F_0(H,\phi)e_{xx}(t)=\frac{\gamma^2\mu_0b_2}{2M_0}(H+M_0)\sin
2\phi e_{xx}(t)$ are both proportional to the strain amplitude
$e_{xx}(t)$. However, their ratio $F_0/\Omega^2_1\propto
(H+M_0)\sin 2\phi/[H+M_0-(3H+2M_0)\cos^2\phi]$, which determines
the relative strength of parametric modulation to driving force,
does not depend on strain and can be adjusted by either changing
the magnitude of the applied field, $H$, and/or the orientation
$\phi$ of the external magnetic field with respect to TG wave
vector.

In figure 1(c), we show the results of the numerical solution of
Eq.~(1). The solution of the equation results in a time, angle,
and field - dependent time trace, which is subsequently Fourier
transformed and displayed as a spectral amplitude. In comparison
to figure 1(b), we note the exceptional similarity between the
experimental data and the calculated response, and in particular
the appearance of harmonics of the underlying elastic waves.  In
the simulation, the SAW and SSLW frequencies are {\it a
posteriori} extracted from the experimental data and the SSLW
amplitude and decay time are calculated by the Green's function
formalism \cite{Janusonis2016_1}. The parametrically driven
upconversion and downconversion are the results of the simulation.

Figure 2(a) shows the dependence of the Fourier spectra in
Fig.~1(b) on the magnetic field at selected frequencies
corresponding to the acoustic SAW, and SSLW, as well as the
nonlinear responses of SSLW-SAW and SAW+SAW frequencies. The
strongest SAW signal displays a resonance at H=250~G. As discussed
previously for this material heterostructure, the largest signal
corresponds to the SAW excitation  which is well approximated by a
Lorentzian line shape (dotted line in the figure).  In Figure 2(b)
we take a closer look at the dependence of SHG signal (SAW+SAW) at
5.10~GHz as a function of the magnetic field, which displays two
pronounced maxima corresponding to the $\Omega_0=\omega$ and
$\Omega_0=2\omega$, both in the experiment and in the numerical
simulation. Here $\omega$ denotes the frequency of surface
acoustic wave. In order to understand the physical origin of this
dependence we have applied first-order perturbation theory to
Eq.~(1) assuming $e_{xx}(t)$ to be a small parameter. Assuming a
monochromatic elastic driving force (SAW only)
$e_{xx}(t)=e_{xx,0}{\rm exp}(i\omega t)$ we obtained the following
analytical expression for the nonlinear correction at frequency
$2\omega$:
\begin{equation}
m(2\omega)\propto\frac{1}{\Omega_0^2-\omega^2+i\omega\Gamma_0}\frac{\Omega^2_1}{\Omega_0^2-4\omega^2+2i\omega\Gamma_0}\,,
\end{equation}
which represents a product of two Lorentzians $L(\omega)$ and
$L(2\omega)$. Therefore, in analogy to nonlinear
optics\cite{TemnovJOPT2016} the spectral dependence of the
second-order susceptibility diplays two resonances at frequencies
$\omega$ and $2\omega$, respectively. However, whereas in
nonlinear optics SHG is caused by  $\chi^{(2)}$-nonlinearities in
the wave equation, in magneto-elastics it originates from the
joint action of the parametric modulation of the FMR frequency and
periodic external driving, both at the same frequency $\omega$.

\begin{figure}
    \centering
    \includegraphics[width=\columnwidth]{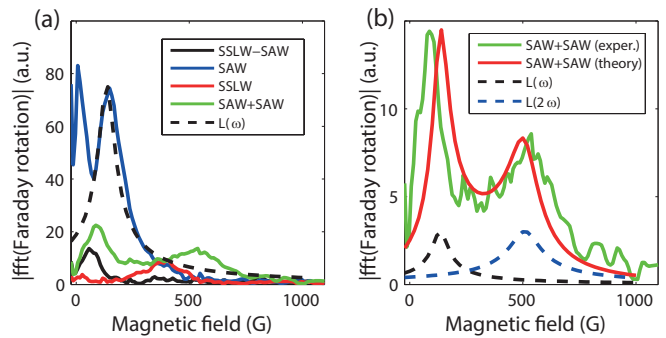}
    \caption{ (a) Horizontal crossections of experimental (Fig.~1(b)) plots corresponding to SAW and SSLW frequencies as well as their difference (SSLW-SAW) and SAW second harmonic (SAW+SAW) frequencies. The dashed line shows the SAW Lorentzian $L(\omega)$. (b) The strongest nonlinear mixing signal (SAW+SAW) is well approximated by an analytical approximation (Eq.~(2)); it displays two maxima corresponding to the resonances $\Omega_0(H=150~{\rm G})=\omega$ and $\Omega_0(H=520~{\rm G})=2\omega$.}
    \label{simulation}
\end{figure}

The above analytical treatment neglects the second driving
frequency at SSLW, which is necessary to explain the sum and
difference frequency wave mixing. In contrast to the long-lived
SAW, the amplitude of SSLW rapidly decays on a time scale of about
1-3~ns, depending on the elastic constants of the substrate, TG
wavelength and nickel thickness.  However, replacing the first
Lorentzian $L(\omega)$ in Eq.~(2) by a sum of two Lorentzians,
$L(\omega)+L(\omega_{SSLW})$, allows to approximate the
experimental data for SSLW-SAW in Fig. 2(a) as well.

\begin{figure}[ht]
    \includegraphics[width=0.8\columnwidth]{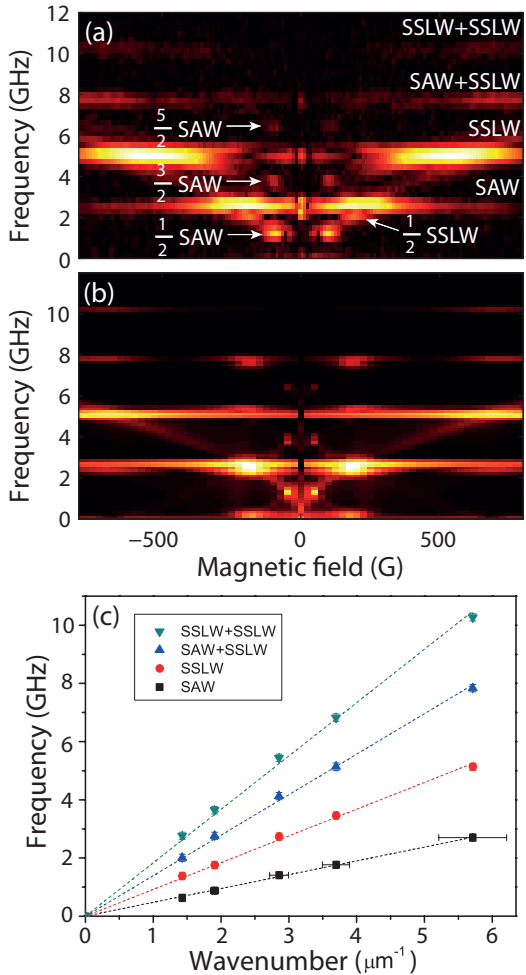}
    \caption{(a) On glass substrates, two strong elastic waves are generated (SAW and SSLW) allowing the observation of sum frequencies at SAW + SSLW (7.83~GHz) and SSLW + SSLW (10.20~GHz) in addition to the linearly activated precession at 2.73~GHz and 5.10~GHz.  Additionally, parametric downconversion is also witnessed at the half frequency of SAW and SSLW (and their integer multiples). (b) Simulations of the results on glass, where higher strain values can be obtained, show excellent agreement with the experimental findings. (c) Dispersion relations for several grating periodicities.  The resultant slope indicates the velocity of the excitation, which for the two lower modes, correspond to the propagation velocity of the underlying elastic waves.  For the sum frequencies, the extracted 'velocities' indicate that higher precessional frequencies always occur at sum of the lower elastic responses.}
   \label{dispersion}
\end{figure}

Engineering the elastic properties of the dielectric substrate can
be used to enhance existing, or create additional, nonlinear
responses. In contrast to the MgO substrate, where the SAW
excitation dominates, a similar nickel/glass structure displays
markedly different behavior due to the strong contribution of the
SSLW excitation\cite{Janusonis2016}, providing additional
opportunities to tune the nonlinear responsivity.  As a
demonstration, we perform the same measurements on the Ni/glass
structure with a 1.1~$\mu$m period, which results in a more
efficient magneto-elastic frequency mixing as shown in Figure
3(a). In addition to the linear SAW and SSLW excitations, there
now exists two responses at $\sim7.83~GHz$ and $\sim10.20~GHz$
which we recognize as the precession response due to the sum
frequencies of the underlying elastic waves, SAW + SSLW and SSLW +
SSLW (Due to the particularities of the elastic properties of the
glass, SAW + SAW excitation overlaps nearly perfectly with the
linear SSLW response and is thus not evident.)  In addition to
these nonlinear sum frequencies, the response on glass is marked
by the appearance of precessional amplitude at the parametrically
downconverted half frequencies (and their multiples) as indicated
in the figure.

Calculations similar to those performed earlier are accomplished
by adjusting the relative strength of SAW and SSLW for the new
configuration.  In this case, the time window over which both SAW
and SSLW act are unchanged, however, the strength of the
excitation is increased for both SAW and SSLW to account for the
more favorable elastic constants of the amorphous glass substrate.
Under these conditions of strong elastic driving, the simulation
is again able to extract the salient features of the experimental
findings, in particular, the generation of parametrically
downconverted frequencies and the sum frequencies. The data and
calculations in Fig. ~3(b) demonstrate the effect for a magnetic
field angle of $\phi=7.5^{\circ}$ (magnetic field nearly collinear
to the TG wavevector), where parametric driving dominates.
Likewise, turning the magnetic field angle to larger values
suppresses these downconverted and upconverted responses.  In the
supplemental we show an example of the response at
$\phi=60^{\circ}$ where the parametric response is suppressed.

Finally, we aggregate the results of several grating periodicities
(1.1, 1.7, 2.2, 3.3 and 4.4~$\mu$m) on the glass substrate, all if
which show sum frequency generation, to show the scaling of
precessional frequency as a function of grating wavenumber. The
extracted velocities for the two lower branches are $3000 \pm
125$~m/s and $5900 \pm 110$~m/s in line with our previous
measurement and published data for the Rayleigh SAW and
longitudinal velocity in glass. For the two upper branches, we
extract values of $8900 \pm 230$~m/s and $11800 \pm 300$~m/s which
overlap, within errors, with the sum frequencies SAW + SSLW and
SSLW + SSLW.  Velocity values are obtained by zero-intercept
linear fits of the dispersion curves, while the horizontal and
vertical error bars on the data account for the uncertainty in
excitation grating period and frequency uncertainty due to the
finite measurement time, respectively.  The dispersion relations
should be understood in the framework of elastic propagation and
the resonant precession that they drive.  While the two lower
branches represent both elastic velocity and precessional
frequency, the upper branches should only be associated with the
precession of magnetization at the sum frequencies of the
underlying elastic waves.


In summary, we have demonstrated the general feature of parametric
excitation of nonlinear magnetization precession under the action
of a driving elastic field. Under an applied field, the
particularities of the response can be tuned to accentuate or
amplify either the sum, difference, or parametric downconverted
frequencies.  The range of nonlinear responses can be further
selected by a careful choice of substrate material, which selects
the relative strengths of the active elastic waves.  To
corroborate our experimental results, we calculate the response of
magnetization under the action of elastic waves to arrive at the
equation for a parametric oscillator.   Numerical and analytical
calculations find excellent qualitative agreement to the data.
This initial demonstration opens the door to more complex elastic
wave\cite{Schulein} control over magnetization that could herald
the emergence of extremely broadband, widely tunable, control of
magnetization precession, including magnetization reorientation,
and elastic activation of quantized magnonic modes.  Furthermore,
the presented methodology can be used to investigate novel
materials with unknown magneto-elastic properties. Thin films of
magnetic MAX-phases with negligible magneto-crystalline anisotropy
and possibly high magneto-strictive coupling could be good
candidates for future research \cite{Salikhov2015}.

Funding from {\it Strat\'{e}gie Internationale} "NNN-Telecom" de
la R\'{e}gion Pays de La Loire, ANR-DFG "PPMI-NANO"
(ANR-15-CE24-0032 \& DFG SE2443/2) and {\it Alexander von Humboldt
Stiftung} is highly appreciated.


\begin{thebibliography}{29}
\expandafter\ifx\csname
natexlab\endcsname\relax\def\natexlab#1{#1}\fi
\expandafter\ifx\csname bibnamefont\endcsname\relax
  \def\bibnamefont#1{#1}\fi
\expandafter\ifx\csname bibfnamefont\endcsname\relax
  \def\bibfnamefont#1{#1}\fi
\expandafter\ifx\csname citenamefont\endcsname\relax
  \def\citenamefont#1{#1}\fi
\expandafter\ifx\csname url\endcsname\relax
  \def\url#1{\texttt{#1}}\fi
\expandafter\ifx\csname
urlprefix\endcsname\relax\def\urlprefix{URL }\fi
\providecommand{\bibinfo}[2]{#2}
\providecommand{\eprint}[2][]{\url{#2}}

\bibitem[{\citenamefont{Butikov}({2004})}]{Butikov}
\bibinfo{author}{\bibfnamefont{E.}~\bibnamefont{Butikov}},
  \bibinfo{journal}{{Eur. J. Phys.}}
  \textbf{\bibinfo{volume}{{25}}}, \bibinfo{pages}{{535}}
  (\bibinfo{year}{{2004}}).

\bibitem[{\citenamefont{Rugar and Grutter}({1991})}]{Rugar}
\bibinfo{author}{\bibfnamefont{D.}~\bibnamefont{Rugar}} \bibnamefont{and}
  \bibinfo{author}{\bibfnamefont{P.}~\bibnamefont{Grutter}},
  \bibinfo{journal}{{Phys. Rev. Lett.}} \textbf{\bibinfo{volume}{{67}}},
  \bibinfo{pages}{{699}} (\bibinfo{year}{{1991}}).

\bibitem[{\citenamefont{Aspelmeyer et~al.}({2014})\citenamefont{Aspelmeyer,
  Kippenberg, and Marquard}}]{Aspelmeyer}
\bibinfo{author}{\bibfnamefont{M.}~\bibnamefont{Aspelmeyer}},
  \bibinfo{author}{\bibfnamefont{T.~J.} \bibnamefont{Kippenberg}},
  \bibnamefont{and} \bibinfo{author}{\bibfnamefont{F.}~\bibnamefont{Marquard}},
  \bibinfo{journal}{{Rev. Mod. Phys.}}
  \textbf{\bibinfo{volume}{{86}}}, \bibinfo{pages}{{1391}}
  (\bibinfo{year}{{2014}}).

\bibitem[{\citenamefont{Papariello et~al.}(2016)\citenamefont{Papariello,
  Zilberberg, Eichler, and Chitra}}]{Papariello2016}
\bibinfo{author}{\bibfnamefont{L.}~\bibnamefont{Papariello}},
  \bibinfo{author}{\bibfnamefont{O.}~\bibnamefont{Zilberberg}},
  \bibinfo{author}{\bibfnamefont{A.}~\bibnamefont{Eichler}}, \bibnamefont{and}
  \bibinfo{author}{\bibfnamefont{R.}~\bibnamefont{Chitra}},
  \bibinfo{journal}{Phys. Rev. E} \textbf{\bibinfo{volume}{94}},
  \bibinfo{pages}{022201} (\bibinfo{year}{2016}).

\bibitem[{\citenamefont{Zhang et~al.}(2002)\citenamefont{Zhang, Baskaran, and
  Turner}}]{ZhangSA2002}
\bibinfo{author}{\bibfnamefont{W.}~\bibnamefont{Zhang}},
  \bibinfo{author}{\bibfnamefont{R.}~\bibnamefont{Baskaran}}, \bibnamefont{and}
  \bibinfo{author}{\bibfnamefont{K.~L.} \bibnamefont{Turner}},
  \bibinfo{journal}{Sens. Actuators A} \textbf{\bibinfo{volume}{102}},
  \bibinfo{pages}{139} (\bibinfo{year}{2002}).

\bibitem[{\citenamefont{L'vov}(1994)}]{L'vov}
\bibinfo{author}{\bibfnamefont{V.~S.} \bibnamefont{L'vov}},
  \emph{\bibinfo{title}{Wave Turbulence Under Parametric Excitation}}, Springer
  Series in Nonlinear Dynamics (\bibinfo{publisher}{Springer Berlin
  Heidelberg}, \bibinfo{year}{1994}), ISBN \bibinfo{isbn}{978-3-642-75297-1}.

\bibitem[{\citenamefont{Castellanos-Beltran
  et~al.}({2008})\citenamefont{Castellanos-Beltran, Irwin, Hilton, Vale, and
  Lehnert}}]{Castellanos-Beltran}
\bibinfo{author}{\bibfnamefont{M.~A.} \bibnamefont{Castellanos-Beltran}},
  \bibinfo{author}{\bibfnamefont{K.~D.} \bibnamefont{Irwin}},
  \bibinfo{author}{\bibfnamefont{G.~C.} \bibnamefont{Hilton}},
  \bibinfo{author}{\bibfnamefont{L.~R.} \bibnamefont{Vale}}, \bibnamefont{and}
  \bibinfo{author}{\bibfnamefont{K.~W.} \bibnamefont{Lehnert}},
  \bibinfo{journal}{{Nature Phys.}} \textbf{\bibinfo{volume}{{4}}},
  \bibinfo{pages}{{929}} (\bibinfo{year}{{2008}}).

\bibitem[{\citenamefont{Jia et~al.}(2016)\citenamefont{Jia, Du, and
  Seshia}}]{JiaSciRep2016}
\bibinfo{author}{\bibfnamefont{Y.}~\bibnamefont{Jia}},
  \bibinfo{author}{\bibfnamefont{S.}~\bibnamefont{Du}}, \bibnamefont{and}
  \bibinfo{author}{\bibfnamefont{A.~A.} \bibnamefont{Seshia}},
  \bibinfo{journal}{Sci. Rep.} \textbf{\bibinfo{volume}{6}},
  \bibinfo{pages}{30167} (\bibinfo{year}{2016}).

\bibitem[{\citenamefont{Zhang et~al.}(2016)\citenamefont{Zhang, Zou, Jiang, and
  Tang}}]{Zhange1501286}
\bibinfo{author}{\bibfnamefont{X.}~\bibnamefont{Zhang}},
  \bibinfo{author}{\bibfnamefont{C.-L.} \bibnamefont{Zou}},
  \bibinfo{author}{\bibfnamefont{L.}~\bibnamefont{Jiang}}, \bibnamefont{and}
  \bibinfo{author}{\bibfnamefont{H.~X.} \bibnamefont{Tang}},
   \bibinfo{journal}{{Sci. Adv.}}
  \textbf{\bibinfo{volume}{2}}, \bibinfo{pages}{1501286}
  (\bibinfo{year}{2016}).

\bibitem[{\citenamefont{Keshtgar et~al.}(2014)\citenamefont{Keshtgar, Zareyan,
  and Bauer}}]{KeshtgarSSC2014}
\bibinfo{author}{\bibfnamefont{H.}~\bibnamefont{Keshtgar}},
  \bibinfo{author}{\bibfnamefont{M.}~\bibnamefont{Zareyan}}, \bibnamefont{and}
  \bibinfo{author}{\bibfnamefont{G.}~\bibnamefont{Bauer}},
  \bibinfo{journal}{Solid State Commun.} \textbf{\bibinfo{volume}{198}},
  \bibinfo{pages}{30} (\bibinfo{year}{2014}).

\bibitem[{\citenamefont{Serga et~al.}(2010)\citenamefont{Serga, Chumak, and
  Hillebrands}}]{Serga}
\bibinfo{author}{\bibfnamefont{A.~A.} \bibnamefont{Serga}},
  \bibinfo{author}{\bibfnamefont{A.~V.} \bibnamefont{Chumak}},
  \bibnamefont{and}
  \bibinfo{author}{\bibfnamefont{B.}~\bibnamefont{Hillebrands}},
  \bibinfo{journal}{J. Phys. D: Appl. Phys.}
  \textbf{\bibinfo{volume}{43}}, \bibinfo{pages}{264002}
  (\bibinfo{year}{2010}).

\bibitem[{\citenamefont{Elezzabi and Irvine}(2003)}]{ElezzabiAPL2003}
\bibinfo{author}{\bibfnamefont{A.~Y.} \bibnamefont{Elezzabi}} \bibnamefont{and}
  \bibinfo{author}{\bibfnamefont{S.~E.} \bibnamefont{Irvine}},
  \bibinfo{journal}{Appl. Phys. Lett.} \textbf{\bibinfo{volume}{82}},
  \bibinfo{pages}{2464} (\bibinfo{year}{2003}).

\bibitem[{\citenamefont{Capua et~al.}(2016)\citenamefont{Capua, Rettner, and
  Parkin}}]{CapuaPRL2016}
\bibinfo{author}{\bibfnamefont{A.}~\bibnamefont{Capua}},
  \bibinfo{author}{\bibfnamefont{C.}~\bibnamefont{Rettner}}, \bibnamefont{and}
  \bibinfo{author}{\bibfnamefont{S.~S.~P.} \bibnamefont{Parkin}},
  \bibinfo{journal}{Phys. Rev. Lett.} \textbf{\bibinfo{volume}{116}},
  \bibinfo{pages}{047204} (\bibinfo{year}{2016}).

\bibitem[{\citenamefont{Gerrits et~al.}(2007)\citenamefont{Gerrits, Krivosik,
  Schneider, Patton, and Silva}}]{GerritsPRL2007}
\bibinfo{author}{\bibfnamefont{T.}~\bibnamefont{Gerrits}},
  \bibinfo{author}{\bibfnamefont{P.}~\bibnamefont{Krivosik}},
  \bibinfo{author}{\bibfnamefont{M.~L.} \bibnamefont{Schneider}},
  \bibinfo{author}{\bibfnamefont{C.~E.} \bibnamefont{Patton}},
  \bibnamefont{and} \bibinfo{author}{\bibfnamefont{T.~J.} \bibnamefont{Silva}},
  \bibinfo{journal}{Phys. Phys. Lett.} \textbf{\bibinfo{volume}{98}},
  \bibinfo{pages}{207602} (\bibinfo{year}{2007}).

\bibitem[{\citenamefont{Bauer et~al.}(2015)\citenamefont{Bauer, Majchrak,
  Kachel, Back, and Woltersdorf}}]{Bauer}
\bibinfo{author}{\bibfnamefont{H.}~\bibnamefont{Bauer}},
  \bibinfo{author}{\bibfnamefont{P.}~\bibnamefont{Majchrak}},
  \bibinfo{author}{\bibfnamefont{T.}~\bibnamefont{Kachel}},
  \bibinfo{author}{\bibfnamefont{C.}~\bibnamefont{Back}}, \bibnamefont{and}
  \bibinfo{author}{\bibfnamefont{G.}~\bibnamefont{Woltersdorf}},
  \bibinfo{journal}{Nat. Commun.} \textbf{\bibinfo{volume}{6}},
  \bibinfo{pages}{8274} (\bibinfo{year}{2015}).

\bibitem[{\citenamefont{Weiler et~al.}({2011})\citenamefont{Weiler, Dreher,
  Heeg, Huebl, Gross, Brandt, and Goennenwein}}]{Weiler2}
\bibinfo{author}{\bibfnamefont{M.}~\bibnamefont{Weiler}},
  \bibinfo{author}{\bibfnamefont{L.}~\bibnamefont{Dreher}},
  \bibinfo{author}{\bibfnamefont{C.}~\bibnamefont{Heeg}},
  \bibinfo{author}{\bibfnamefont{H.}~\bibnamefont{Huebl}},
  \bibinfo{author}{\bibfnamefont{R.}~\bibnamefont{Gross}},
  \bibinfo{author}{\bibfnamefont{M.~S.} \bibnamefont{Brandt}},
  \bibnamefont{and} \bibinfo{author}{\bibfnamefont{S.~T.~B.}
  \bibnamefont{Goennenwein}}, \bibinfo{journal}{{Phys. Rev. Lett.}}
  \textbf{\bibinfo{volume}{{106}}}, \bibinfo{pages}{{117601}}
  (\bibinfo{year}{{2011}}).

\bibitem[{\citenamefont{Thevenard et~al.}({2014})\citenamefont{Thevenard,
  Gourdon, Prieur, von Bardeleben, Vincent, Becerra, Largeau, and
  Duquesne}}]{Thevenard2}
\bibinfo{author}{\bibfnamefont{L.}~\bibnamefont{Thevenard}},
  \bibinfo{author}{\bibfnamefont{C.}~\bibnamefont{Gourdon}},
  \bibinfo{author}{\bibfnamefont{J.~Y.} \bibnamefont{Prieur}},
  \bibinfo{author}{\bibfnamefont{H.~J.} \bibnamefont{von Bardeleben}},
  \bibinfo{author}{\bibfnamefont{S.}~\bibnamefont{Vincent}},
  \bibinfo{author}{\bibfnamefont{L.}~\bibnamefont{Becerra}},
  \bibinfo{author}{\bibfnamefont{L.}~\bibnamefont{Largeau}}, \bibnamefont{and}
  \bibinfo{author}{\bibfnamefont{J.~Y.} \bibnamefont{Duquesne}},
  \bibinfo{journal}{{Phys. Rev. B}} \textbf{\bibinfo{volume}{{90}}},
  \bibinfo{pages}{{094401}} (\bibinfo{year}{{2014}}).

\bibitem[{\citenamefont{Dreher et~al.}({2012})\citenamefont{Dreher, Weiler,
  Pernpeintner, Huebl, Gross, Brandt, and Goennenwein}}]{Dreher}
\bibinfo{author}{\bibfnamefont{L.}~\bibnamefont{Dreher}},
  \bibinfo{author}{\bibfnamefont{M.}~\bibnamefont{Weiler}},
  \bibinfo{author}{\bibfnamefont{M.}~\bibnamefont{Pernpeintner}},
  \bibinfo{author}{\bibfnamefont{H.}~\bibnamefont{Huebl}},
  \bibinfo{author}{\bibfnamefont{R.}~\bibnamefont{Gross}},
  \bibinfo{author}{\bibfnamefont{M.~S.} \bibnamefont{Brandt}},
  \bibnamefont{and} \bibinfo{author}{\bibfnamefont{S.~T.~B.}
  \bibnamefont{Goennenwein}}, \bibinfo{journal}{{Phys. Rev. B}}
  \textbf{\bibinfo{volume}{{86}}}, \bibinfo{pages}{{134415}}
  (\bibinfo{year}{{2012}}).

\bibitem[{\citenamefont{Kim et~al.}({2012})\citenamefont{Kim, Vomir, and
  Bigot}}]{Kim}
\bibinfo{author}{\bibfnamefont{J.-W.} \bibnamefont{Kim}},
  \bibinfo{author}{\bibfnamefont{M.}~\bibnamefont{Vomir}}, \bibnamefont{and}
  \bibinfo{author}{\bibfnamefont{J.-Y.} \bibnamefont{Bigot}},
  \bibinfo{journal}{{Phys. Rev. Lett.}} \textbf{\bibinfo{volume}{{109}}},
  \bibinfo{pages}{{166601}} (\bibinfo{year}{{2012}}).

\bibitem[{\citenamefont{Scherbakov et~al.}({2010})\citenamefont{Scherbakov,
  Salasyuk, Akimov, Liu, Bombeck, Brueggemann, Yakovlev, Sapega, Furdyna, and
  Bayer}}]{Scherbakov}
\bibinfo{author}{\bibfnamefont{A.~V.} \bibnamefont{Scherbakov}},
  \bibinfo{author}{\bibfnamefont{A.~S.} \bibnamefont{Salasyuk}},
  \bibinfo{author}{\bibfnamefont{A.~V.} \bibnamefont{Akimov}},
  \bibinfo{author}{\bibfnamefont{X.}~\bibnamefont{Liu}},
  \bibinfo{author}{\bibfnamefont{M.}~\bibnamefont{Bombeck}},
  \bibinfo{author}{\bibfnamefont{C.}~\bibnamefont{Brueggemann}},
  \bibinfo{author}{\bibfnamefont{D.~R.} \bibnamefont{Yakovlev}},
  \bibinfo{author}{\bibfnamefont{V.~F.} \bibnamefont{Sapega}},
  \bibinfo{author}{\bibfnamefont{J.~K.} \bibnamefont{Furdyna}},
  \bibnamefont{and} \bibinfo{author}{\bibfnamefont{M.}~\bibnamefont{Bayer}},
  \bibinfo{journal}{{Phys. Rev. Lett.}} \textbf{\bibinfo{volume}{{105}}},
  \bibinfo{pages}{{117204}} (\bibinfo{year}{{2010}}).

\bibitem[{\citenamefont{Jaeger et~al.}({2013})\citenamefont{Jaeger, Scherbakov,
  Linnik, Yakovlev, Wang, Wadley, Holy, Cavill, Akimov, Rushforth
  et~al.}}]{Jaeger1}
\bibinfo{author}{\bibfnamefont{J.~V.} \bibnamefont{Jaeger}},
  \bibinfo{author}{\bibfnamefont{A.~V.} \bibnamefont{Scherbakov}},
  \bibinfo{author}{\bibfnamefont{T.~L.} \bibnamefont{Linnik}},
  \bibinfo{author}{\bibfnamefont{D.~R.} \bibnamefont{Yakovlev}},
  \bibinfo{author}{\bibfnamefont{M.}~\bibnamefont{Wang}},
  \bibinfo{author}{\bibfnamefont{P.}~\bibnamefont{Wadley}},
  \bibinfo{author}{\bibfnamefont{V.}~\bibnamefont{Holy}},
  \bibinfo{author}{\bibfnamefont{S.~A.} \bibnamefont{Cavill}},
  \bibinfo{author}{\bibfnamefont{A.~V.} \bibnamefont{Akimov}},
  \bibinfo{author}{\bibfnamefont{A.~W.} \bibnamefont{Rushforth}},
  \bibnamefont{et~al.}, \bibinfo{journal}{{Appl. Phys. Lett.}}
  \textbf{\bibinfo{volume}{{103}}}, \bibinfo{pages}{{032409}}
  (\bibinfo{year}{{2013}}).

\bibitem[{\citenamefont{Afanasiev et~al.}({2014})\citenamefont{Afanasiev,
  Razdolski, Skibinsky, Bolotin, Yagupov, Strugatsky, Kirilyuk, Rasing, and
  Kimel}}]{Afanasiev}
\bibinfo{author}{\bibfnamefont{D.}~\bibnamefont{Afanasiev}},
  \bibinfo{author}{\bibfnamefont{I.}~\bibnamefont{Razdolski}},
  \bibinfo{author}{\bibfnamefont{K.~M.} \bibnamefont{Skibinsky}},
  \bibinfo{author}{\bibfnamefont{D.}~\bibnamefont{Bolotin}},
  \bibinfo{author}{\bibfnamefont{S.~V.} \bibnamefont{Yagupov}},
  \bibinfo{author}{\bibfnamefont{M.~B.} \bibnamefont{Strugatsky}},
  \bibinfo{author}{\bibfnamefont{A.}~\bibnamefont{Kirilyuk}},
  \bibinfo{author}{\bibfnamefont{T.}~\bibnamefont{Rasing}}, \bibnamefont{and}
  \bibinfo{author}{\bibfnamefont{A.~V.} \bibnamefont{Kimel}},
  \bibinfo{journal}{{Phys. Rev. Lett.}} \textbf{\bibinfo{volume}{{112}}}
  (\bibinfo{year}{{2014}}).

\bibitem[{\citenamefont{Janu{\v{s}}onis
  et~al.}(2016)\citenamefont{Janu{\v{s}}onis, Jansma, Chang, Q., Gatilova,
  Lomonosov, Shalagatskyi, Pezeril, Temnov, and Tobey}}]{Janusonis2016_1}
\bibinfo{author}{\bibfnamefont{J.}~\bibnamefont{Janu{\v{s}}onis}},
  \bibinfo{author}{\bibfnamefont{T.}~\bibnamefont{Jansma}},
  \bibinfo{author}{\bibfnamefont{C.~L.} \bibnamefont{Chang}},
  \bibinfo{author}{\bibfnamefont{L.}~\bibnamefont{Q.}},
  \bibinfo{author}{\bibfnamefont{A.}~\bibnamefont{Gatilova}},
  \bibinfo{author}{\bibfnamefont{A.~M.} \bibnamefont{Lomonosov}},
  \bibinfo{author}{\bibfnamefont{V.}~\bibnamefont{Shalagatskyi}},
  \bibinfo{author}{\bibfnamefont{T.}~\bibnamefont{Pezeril}},
  \bibinfo{author}{\bibfnamefont{V.~V.} \bibnamefont{Temnov}},
  \bibnamefont{and} \bibinfo{author}{\bibfnamefont{R.~I.} \bibnamefont{Tobey}},
  \bibinfo{journal}{Sci. Rep.} \textbf{\bibinfo{volume}{6}},
  \bibinfo{pages}{29143} (\bibinfo{year}{2016}).

\bibitem[{\citenamefont{Janu{\v{s}}onis
  et~al.}(2015)\citenamefont{Janu{\v{s}}onis, Chang, van Loosdrecht, and
  Tobey}}]{Janusonis2015}
\bibinfo{author}{\bibfnamefont{J.}~\bibnamefont{Janu{\v{s}}onis}},
  \bibinfo{author}{\bibfnamefont{C.~L.} \bibnamefont{Chang}},
  \bibinfo{author}{\bibfnamefont{P.~H.~M.} \bibnamefont{van Loosdrecht}},
  \bibnamefont{and} \bibinfo{author}{\bibfnamefont{R.~I.} \bibnamefont{Tobey}},
  \bibinfo{journal}{App. Phys. Let.} \textbf{\bibinfo{volume}{106}},
  \bibinfo{pages}{181601} (\bibinfo{year}{2015}).

\bibitem[{\citenamefont{Janu\ifmmode~\check{s}\else \v{s}\fi{}onis
  et~al.}(2016)\citenamefont{Janu\ifmmode~\check{s}\else \v{s}\fi{}onis, Chang,
  Jansma, Gatilova, Vlasov, Lomonosov, Temnov, and Tobey}}]{Janusonis2016}
\bibinfo{author}{\bibfnamefont{J.}~\bibnamefont{Janu\ifmmode~\check{s}\else
  \v{s}\fi{}onis}}, \bibinfo{author}{\bibfnamefont{C.~L.} \bibnamefont{Chang}},
  \bibinfo{author}{\bibfnamefont{T.}~\bibnamefont{Jansma}},
  \bibinfo{author}{\bibfnamefont{A.}~\bibnamefont{Gatilova}},
  \bibinfo{author}{\bibfnamefont{V.~S.} \bibnamefont{Vlasov}},
  \bibinfo{author}{\bibfnamefont{A.~M.} \bibnamefont{Lomonosov}},
  \bibinfo{author}{\bibfnamefont{V.~V.} \bibnamefont{Temnov}},
  \bibnamefont{and} \bibinfo{author}{\bibfnamefont{R.~I.} \bibnamefont{Tobey}},
  \bibinfo{journal}{Phys. Rev. B} \textbf{\bibinfo{volume}{94}},
  \bibinfo{pages}{024415} (\bibinfo{year}{2016}).

\bibitem[{\citenamefont{Kovalenko et~al.}({2013})\citenamefont{Kovalenko,
  Pezeril, and Temnov}}]{Kovalenko}
\bibinfo{author}{\bibfnamefont{O.}~\bibnamefont{Kovalenko}},
  \bibinfo{author}{\bibfnamefont{T.}~\bibnamefont{Pezeril}}, \bibnamefont{and}
  \bibinfo{author}{\bibfnamefont{V.~V.} \bibnamefont{Temnov}},
  \bibinfo{journal}{{Phys. Rev. Lett.}} \textbf{\bibinfo{volume}{{110}}},
  \bibinfo{pages}{{266602}} (\bibinfo{year}{{2013}}).

\bibitem[{\citenamefont{Temnov et~al.}(2016)\citenamefont{Temnov, Razdolski,
  Pezeril, Makarov, Seletskiy, Melnikov, and Nelson}}]{TemnovJOPT2016}
\bibinfo{author}{\bibfnamefont{V.~V.} \bibnamefont{Temnov}},
  \bibinfo{author}{\bibfnamefont{I.}~\bibnamefont{Razdolski}},
  \bibinfo{author}{\bibfnamefont{T.}~\bibnamefont{Pezeril}},
  \bibinfo{author}{\bibfnamefont{D.}~\bibnamefont{Makarov}},
  \bibinfo{author}{\bibfnamefont{D.}~\bibnamefont{Seletskiy}},
  \bibinfo{author}{\bibfnamefont{A.}~\bibnamefont{Melnikov}}, \bibnamefont{and}
  \bibinfo{author}{\bibfnamefont{K.~A.} \bibnamefont{Nelson}},
  \bibinfo{journal}{J. Opt.} \textbf{\bibinfo{volume}{18}},
  \bibinfo{pages}{093002} (\bibinfo{year}{2016}).

\bibitem[{\citenamefont{Schulein et~al.}(2015)\citenamefont{Schulein, Zallo,
  Atkinson, Schmidt, Trotta, Rastelli, Wixforth, and Krenner}}]{Schulein}
\bibinfo{author}{\bibfnamefont{F.~J.~R.} \bibnamefont{Schulein}},
  \bibinfo{author}{\bibfnamefont{E.}~\bibnamefont{Zallo}},
  \bibinfo{author}{\bibfnamefont{P.}~\bibnamefont{Atkinson}},
  \bibinfo{author}{\bibfnamefont{O.~G.} \bibnamefont{Schmidt}},
  \bibinfo{author}{\bibfnamefont{R.}~\bibnamefont{Trotta}},
  \bibinfo{author}{\bibfnamefont{A.}~\bibnamefont{Rastelli}},
  \bibinfo{author}{\bibfnamefont{A.}~\bibnamefont{Wixforth}}, \bibnamefont{and}
  \bibinfo{author}{\bibfnamefont{H.~J.} \bibnamefont{Krenner}},
  \bibinfo{journal}{Nat. Nano.} \textbf{\bibinfo{volume}{10}},
  \bibinfo{pages}{512} (\bibinfo{year}{2015}).

\bibitem[{\citenamefont{Salikhov et~al.}(2015)\citenamefont{Salikhov,
  Semisalova, Petruhins, Ingason, Rosen, Wiedwald, and Farle}}]{Salikhov2015}
\bibinfo{author}{\bibfnamefont{R.}~\bibnamefont{Salikhov}},
  \bibinfo{author}{\bibfnamefont{A.}~\bibnamefont{Semisalova}},
  \bibinfo{author}{\bibfnamefont{A.}~\bibnamefont{Petruhins}},
  \bibinfo{author}{\bibfnamefont{A.}~\bibnamefont{Ingason}},
  \bibinfo{author}{\bibfnamefont{J.}~\bibnamefont{Rosen}},
  \bibinfo{author}{\bibfnamefont{U.}~\bibnamefont{Wiedwald}}, \bibnamefont{and}
  \bibinfo{author}{\bibfnamefont{M.}~\bibnamefont{Farle}},
  \bibinfo{journal}{Mat. Res. Lett.} \textbf{\bibinfo{volume}{3}},
  \bibinfo{pages}{156} (\bibinfo{year}{2015}).

\end{thebibliography}
\end{document}